\documentclass[a4paper,11pt]{article}
\pdfoutput=1 
\usepackage{jheppub}
\usepackage[T1]{fontenc} \usepackage{comment}
\usepackage{amsmath, amsthm, amssymb, calrsfs, wasysym, verbatim, bbm, graphics, cancel}
\usepackage{tcolorbox}
\usepackage{mathrsfs}
\usepackage{physics}
\usepackage{slashed}
\usepackage{enumitem}
\usepackage{microtype}
\usepackage{hyperref}
\usepackage[nameinlink,noabbrev]{cleveref}
\usepackage{xcolor}
\usepackage[normalem]{ulem}

\numberwithin{equation}{section}

\newcommand{\AdS}{\mathrm{AdS}}

\newcommand{\SL}{\mathrm{SL}}
\newcommand{\Hol}{\mathrm{Hol}}
\newcommand{\R}{\mathbbm R}

\newcommand{\Diff}{\mathrm{Diff}}

\newcommand{\CS}{\mathrm{CS}}

\newcommand{\del}{\partial}
\newcommand{\be}{\begin{equation}}
\newcommand{\ee}{\end{equation}}

\newcommand{\eqn}[1]{(\ref{#1})}

\title{\boldmath Dimensional reduction of $\AdS_3$ Chern-Simons gravity: Schwarzian and affine boundary theories }

\author[a,b]{G.~Chirco,}
\author[a,b]{L.~Vacchiano}
\author[a,b,c]{and P.~Vitale}

\affiliation[a]{Dipartimento di Fisica Ettore Pancini, Universit\`a degli Studi di Napoli Federico II,
Via Cintia 80126,  Napoli, Italy}
\affiliation[b]{INFN, Sezione di Napoli, Italy}
\affiliation[c]{School of Theoretical Physics Dublin Institute for Advanced Studies, 10 Burlington Road, Dublin 4, Ireland}

\emailAdd{goffredo.chirco@unina.it}
\emailAdd{lucio.vacchiano@unina.it}
\emailAdd{patrizia.vitale@unina.it}

\begin{document}

\abstract{We study a symmetry-reduced sector of $AdS_3$ gravity formulated as an $SO(2,2)$ Chern--Simons theory on a three-dimensional manifold with toroidal boundary. The reduction is implemented by requiring a global symmetry and restricting to the sector where the gauge connection is invariant along the symmetry flow. The resulting theory reduces to a two-dimensional BF-like model together with an induced one-dimensional boundary action, whose form is fixed by the three-dimensional Chern--Simons origin. On the boundary subspace selected by suitable mixed boundary conditions, the one-dimensional action reproduces the standard Drinfel'd--Sokolov restriction to coadjoint orbits of the Virasoro group, among them the Schwarzian dynamics associated with JT gravity. Moreover,  the $\mathfrak{so}(2,2)$ algebra of the three-dimensional Chern--Simons model naturally generates  current-dressed Kac--Moody extensions of the one-dimensional boundary dynamics. The full one dimensional boundary dynamics of the reduced theory is then shown to be compatible SYK-like tensor models with global symmetries.}

\newcommand{\bb}[1]{\textcolor{blue}{#1}}
\maketitle

\flushbottom

\section{Introduction}

The correspondence between the Jackiw--Teitelboim (JT) gravity and the Sachdev--Ye--Kitaev (SYK) model 
 provides one of the simplest examples of holographic duality, where a
topological gravitational bulk theory gives rise to non-trivial
one-dimensional boundary dynamics governed by the Schwarzian action
\cite{Jackiw:1984je,Teitelboim:1983ux,Almheiri:2014cka,Maldacena:2016upp,Engelsoy:2016xyb,Cvetic:2016eiv,Mertens:2018fds,Mertens:2022mtm}.
In the gauge-theoretic formulation, JT gravity is described by an
$\mathrm{SL}(2,\mathbb R)$ BF theory, and the Schwarzian theory emerges
after imposing suitable asymptotic and integrability conditions and
performing a Drinfel'd--Sokolov reduction
\cite{Isler:1989hq,Cangemi:1992bj,Ikeda:1993fh,Schaller:1994es,Valach:2019arb,Blommaert:2018oro}. This framework has been generalized in several directions, including extensions with additional current sectors and deformations relevant to non-extremal or near-horizon regimes \cite{Witten:2016iux,Gross:2016kjj,Yoon:2017nig,Liu:2019niv,Carrozza:2018psc,Carlip:2022jfj}. 
A natural question is whether these two-dimensional constructions of generalized JT models can
themselves be obtained from a higher-dimensional gravitational theory in such
a way that both the bulk BF structure and the associated boundary
dynamics follow directly from dimensional reduction. We are particularly interested in obtaining the current extended version of JT gravity in \cite{Chirco:2024ubu, Gonzalez:2018enk} directly from the dimensional reduction of the 3D gravitational action, providing then a well defined gravitational dual for generalized SYK models as in \cite{Yoon:2017nig, Liu:2019niv,Gross:2016kjj, Narayan:2023wlk}. 
Three-dimensional
gravity with negative cosmological constant provides a particularly
natural setting for this problem, since it admits a formulation as an
$\mathrm{SO}(2,2)$ Chern--Simons theory, with the standard Lie algebra decomposition
\begin{equation}
    \mathfrak{so}(2,2)
    \simeq
    \mathfrak{sl}(2,\mathbb R)\oplus
    \mathfrak{sl}(2,\mathbb R),
\end{equation}
and, like BF theory, it has no local propagating bulk degrees of
freedom. Its physical content is instead encoded in global data,
holonomies, boundary conditions and edge degrees of freedom
\cite{Achucarro:1986vz,Witten:1988hc,Carlip:1998qv,Witten:2007kt}.

In this work we study a symmetry-reduced sector of Euclidean
$\mathrm{AdS}_3$ Chern--Simons gravity on a solid torus.
We choose the Euclidean-time cycle to be contractible in the bulk and
the angular cycle to be non-contractible, as appropriate to the
Euclidean BTZ geometry. We then impose invariance of the gauge
connection $(A,\bar{A})$ along the non-contractible angular direction and perform a
dimensional reduction along this circle. The reduced connection
decomposes into  two-dimensional gauge fields $(A_\perp,\bar{A}_{\perp})$ and  adjoint
scalars $(\Phi, \bar{\Phi})$, so that the flatness equations reduce to
\begin{equation}
    F_\perp=0,
    \qquad
    D_\perp\Phi=0,
\end{equation}
\begin{equation}
    \bar{F}_\perp=0,
    \qquad
    D_\perp\bar{\Phi}=0,
\end{equation}
namely the equations of motion of two uncoupled, BF-like theories in two dimensions.
 For each  $\operatorname{SL}(2,\mathbb{R})$ sector, an important feature of the reduction is that the three-dimensional
Chern--Simons action does not reduce to the bare BF bulk action alone.
Rather, one obtains
\begin{equation}
\label{redactionCS}
    S_{\rm red}
    = k_{CS}\int_{M_2}\operatorname{Tr}(\Phi F_\perp)
    -
    \frac{k_{CS}}{2}
    \int_{\partial M_2}
    \operatorname{Tr}(\Phi A_\tau)\,d\tau ,
\end{equation}
and analogously for the other sector, 
so that a definite one-dimensional boundary term is inherited directly
from the parent Chern--Simons theory. This observation is central to
our analysis. In standard BF formulations of JT gravity, a boundary
term is instead usually introduced in order to render the variational
principle well defined once a suitable integrability condition relating
the boundary connection and the dilaton is imposed. Here the boundary
functional is not chosen independently: it is fixed by the
three-dimensional origin of the theory.
This distinction becomes relevant when considering mixed boundary
conditions.

In the JT/BF literature, one may formulate the relation
between the boundary connection and the adjoint scalar either in a
non-covariant form or in a gauge-covariant completion involving a
boundary group-valued field. In the present  setting, focusing on a single $SL(2,\R)$ sector,  the two prescriptions amount to choose
\begin{equation}
    A_\tau=\lambda'(\tau)\Phi ,
\end{equation}
and
\begin{equation}
    A_\tau=
    \lambda'(\tau)\Phi
    +
    u^{-1}\partial_\tau u 
\end{equation}
 where $\lambda(\tau)$ is an orientation-preserving diffeomorphism of the boundary, with $\lambda'>0$  and $u:\del M_2\rightarrow SL(2,\R)$.  
In a purely two-dimensional BF analysis the second expression can be
understood as a gauge-covariant integrability condition and leads to
the same Schwarzian boundary dynamics as the first one, after the appropriate boundary
term has been constructed. In two dimensions, covariant and non covariant conditions lead to the same counter-term needed to absorb the spurious boundary variation of the bulk BF theory.
In three dimensions, although leading to different action functionals at the boundary, they are expected to be equivalent up to reparametrisation. In the remainder of the paper we focus on the boundary dynamics associated with the non-covariant prescription. After imposing the
Drinfel'd--Sokolov restriction, the latter leads to the standard Virasoro coadjoint
transformation
\begin{equation}
    \mathcal L_\lambda
    =
    -\frac{C}{2}\lambda'^2
    +
    \frac12\{\lambda,\tau\}_S,
\end{equation}
where $\{\cdot, \cdot\}_S$ indicates the Schwarzian derivative and 
\begin{equation}
    C=\operatorname{Tr}(\Phi^2)
\end{equation}
is the quadratic Casimir of the reduced BF theory. The on-shell
action \eqn{redactionCS} is proportional to the latter, which is  conserved, but, as in the
standard JT analysis, can be re-expressed as a functional of
the boundary reparametrization mode. In this way the Chern--Simons
reduction reproduces the Schwarzian boundary dynamics for a single $\SL(2,\R)$ sector while, as we will see later, the second sector provides additional degrees of freedom which couple to the reparameterization mode $\lambda$ after suitable boundary conditions are imposed.
The global information inherited from the three-dimensional geometry
plays an essential role in selecting the relevant Virasoro orbit.
Smoothness along the contractible Euclidean-time cycle requires the
corresponding holonomy to lie in the center of the gauge group. For a
thermal circle of length $\beta$, this condition fixes the constant
Drinfel'd--Sokolov representative to
\begin{equation}
    \mathcal L_0^{(n)}
    =
    \frac{\pi^2 n^2}{\beta^2}, \quad n \neq 0
\end{equation}
and hence selects the exceptional Virasoro orbits $\mathcal{O_n}$ given by
\begin{equation}
  \mathcal{O}_n=  \frac{\Diff(S^1)}{\mathrm{SL}^{(n)}(2,\mathbb R)} .
\end{equation}
The fundamental sector $n=1$ gives the thermal Schwarzian orbit
relevant for JT gravity.
We then return to the full
$\mathfrak{so}(2,2)\simeq
\mathfrak{sl}(2,\mathbb R)\oplus\mathfrak{sl}(2,\mathbb R)$
structure. Treating the second sector as an independent copy gives a
decoupled one-dimensional system. A more interesting possibility is to
organize it as a current sector acted upon by the reparametrization
mode of the gravitational sector. This leads to a
Virasoro--Kac--Moody semidirect-product structure and to
a current-dressed Schwarzian action of the type relevant for SYK-like
models with global symmetries
\cite{Gonzalez:2018enk,Chirco:2024so22,Yoon:2017nig,Gross:2016kjj,Liu:2019niv,Narayan:2023wlk,Carrozza:2018psc}.

The paper is organized as follows.
In Section~\ref{sec:3dgravity} we review the Chern--Simons formulation of
three-dimensional $AdS$ gravity and specify the global solid-torus
geometry relevant for the reduction.
In Section~\ref{sec:reduction} we impose the symmetry constraint along the
non-contractible angular direction and derive the reduced BF-like
equations.
In Section~\ref{sec:actionreduction} we perform the reduction at the level of the action and
identify the one-dimensional boundary term inherited from the
Chern--Simons theory.
Section~\ref{sec:vp} discusses the variational principle and the  boundary
conditions.
In Section~\ref{sec:deformed} we implement the Drinfel'd--Sokolov reduction and relate
the reduced Casimir, the thermal holonomy and the exceptional Virasoro
orbits underlying the Schwarzian sector.
In Section~\ref{sec:KMextensions} we discuss the second $\mathrm{SL}(2,\mathbb R)$ sector
and the emergence of current-dressed Virasoro/Kac--Moody boundary
dynamics.
We conclude in Section~~\ref{discussion} with a discussion of the relation between the
three-dimensional Chern--Simons origin of the boundary action and the
standard two-dimensional BF construction of JT gravity.

\section{Three-dimensional gravity as a \texorpdfstring{$AdS_3$\;}{AdS3} Chern--Simons theory}
\label{sec:3dgravity}

Three-dimensional Einstein pure gravity  admits a particularly simple gauge-theoretic formulation as a three-dimensional Chern-Simons theory based on the gauge group $SO(2,2)$ for $\Lambda<0$, $ISO(2,1)$
for $\Lambda=0$, or $SO(3, 1)$ for $\Lambda>0$, with $\Lambda$ the cosmological constant.~\cite{Achucarro:1986vz,Witten:1988hc,Carlip:1998qv,Witten:2007kt}.
The basic reason is that, as in two dimensions, the dreibeine and spin connection can be combined into a single gauge field and, moreover,  the gravitational Lagrangian density  can be recast as  Chern-Simons three form \cite{Witten:1988hc}. We consider $\Lambda<0$, since we are interested in the reduction from three--dimensional to two--dimensional AdS spaces with boundary. 
To this, let $M_3$ be an oriented three-manifold with boundary, and let us consider a Chern--Simons theory with gauge group $SO(2,2)$. The Lie algebra generators,  $J_a, P_a, \, a=1,\dots 3$ obey the commutation relations
\begin{equation}
[J_a,J_b]=\epsilon_{ab}{}^{c}J_c,
\qquad
[J_a,P_b]=\epsilon_{ab}{}^{c}P_c,
\qquad
[P_a,P_b]=\frac{1}{\ell^2}\epsilon_{ab}{}^{c}J_c,
\label{eq:adsalgebrasec2}
\end{equation}
with $\ell$ a real parameter. 
This is referred to as the gravitational basis after the introduction of the gauge connection 
\begin{equation}
\Omega=\omega^a J_a + e^a P_a,
\label{eq:omegagravsec2}
\end{equation}
and the identification of 
$e^a$ as  the dreibeine, $\omega^a$ as  the  spin connection, and $\ell$ the AdS radius, which is related to the cosmological constant through the relation $\Lambda = -1/\ell^2$.
$J_a$, $P_a$ are respectively the Lorentz and AdS--translations generators. The CS action
\be
S[\Omega]=\int_{M_3} \langle\Omega\wedge d \Omega+ \frac{2}{3} \Omega\wedge \Omega\wedge \Omega\rangle
\ee
can be seen to reproduce the Einstein-Hilbert action with negative cosmological constant up to an overall constant,  after associating  the bilinear form $\langle\cdot,\cdot\rangle$ in the Lie algebra  with the invariant  metric
\be
\langle J_a, J_b\rangle = \langle P_a, P_b\rangle = 0,\;\;\; \langle J_a, P_b\rangle = \delta_{ab}.
\ee
At the Lie-algebra level one has the  isomorphism
\begin{equation}
\mathfrak{so}(2,2)\simeq \mathfrak{sl}(2,\mathbb R)\oplus \mathfrak{sl}(2,\mathbb R),
\label{eq:so22splitsec2}
\end{equation}
so that the $\mathfrak{so}(2,2)$-valued connection may be decomposed into two commuting $\mathfrak{sl}(2,\mathbb R)$ connections,
\begin{equation}
\Omega=(A,\bar A).
\label{eq:chiralomegasec2}
\end{equation}
Indeed, the two commuting algebras are spanned by  the  generators
\begin{equation}
J_a^\pm=\frac12\bigl(J_a\pm \ell P_a\bigr),
\label{eq:chiralgeneratorssec2}
\end{equation}
and $A, \bar A$  are given by 
\begin{equation}
A=\left(\omega^a+\frac{1}{\ell}e^a\right)J_a^+,
\qquad
\bar A=\left(\omega^a-\frac{1}{\ell}e^a\right)J_a^-.
\label{eq:chiralconnectionssec2}
\end{equation}
Therefore, the Einstein--Hilbert action with negative cosmological constant is  equivalent, up to boundary terms, to the difference of two $SL(2,\R)$--Chern--Simons actions,
\begin{equation}
S_{\mathrm{grav}}
=
S_{\CS}[A]-S_{\CS}[\bar A],
\label{eq:gravascssec2}
\end{equation}
where, restoring the appropriate constants,
\begin{equation}\label{eq:csactionsec2}
S_{\CS}[A]
=
\frac{k_{CS}}{4\pi}\int_{M_3} \operatorname{Tr} \bigl(A\wedge \dd A+ \frac{2}{3} A\wedge A\wedge A\bigr)
\ee
and similarly for $\bar A$. The Chern--Simons level is fixed by the gravitational coupling to be
\begin{equation}
k_{CS}=\frac{\ell}{4G}.
\label{eq:levelsec2}
\end{equation}
The equations of motion are simply the flatness conditions
\begin{equation}
F(A)=0,
\qquad
F(\bar A)=0.
\label{eq:flateomsec2}
\end{equation}
In gravitational variables these are equivalent to the vanishing of the torsion and to the Einstein equations with negative cosmological constant,
\begin{equation}
T^a=de^a+\epsilon^{a}{}_{bc}\,\omega^b\wedge e^c=0,
\label{eq:torsionsec2}
\end{equation}
\begin{equation}
R^a+\frac{1}{2\ell^2}\epsilon^{a}{}_{bc}\,e^b\wedge e^c=0.
\label{eq:einsteinsec2}
\end{equation}
Thus the Chern--Simons formulation captures three-dimensional gravity as a flat gauge theory whose non-trivial content is encoded globally, through holonomies and boundary data, rather than through local propagating bulk degrees of freedom~\cite{Witten:1988hc,Carlip:1998qv,Witten:2007kt}.
  For our purpose, we further specialize to the case
\begin{equation}
M_3=M_2\times S^1,
\label{eq:M3productsec2}
\end{equation}
where $M_2$ is topologically a disk, with boundary
$\partial M_2=S^1
$, and the boundary of the three-manifold is a two-torus,
\begin{equation}
\partial M_3=T^2.
\label{eq:T2boundarysec2}
\end{equation}
This is the natural setting for the reduction we want to perform: one circle will play the role of the direction along which we impose invariance, while the remaining boundary circle will support the induced one-dimensional dynamics.
In particular, we denote by 
\begin{equation} \label{coordinates}
x^\mu=(\rho,\tau,\phi),
\end{equation}
the local coordinates of $M_3$,\
where $\rho$ is the radial coordinate on the disk, and $(\tau,\phi)$ the  coordinates on the boundary torus. We shall interpret $\tau$ as the Euclidean time and we take the $\tau$-circle to be contractible \cite{Banados:1992wn,Maloney:2007ud}. Correspondingly, the  
angular coordinate $\phi$ runs along the non-contractible circle. This specifies to an Euclidean BTZ geometry in the bulk. The $M_2$ disk is therefore parametrized by the polar coordinates
\begin{equation}
x_\perp=(\rho,\tau)\,.
\end{equation}
  In particular, Euclidean BTZ geometries are locally $AdS_3$, but differ globally by the holonomies of the flat  connections around the two cycles of the boundary torus \cite{Banados:1992wn,Carlip:1998qv}. The relation between three-dimensional gravity, asymptotic boundary dynamics and induced boundary degrees of freedom has been extensively studied \cite{Brown:1986nw,Coussaert:1995zp,Arcioni:2002vv}.
In the next section, we implement a dimensional reduction of the model by restricting to connections invariant along the flow of the globally defined vector field $X=\partial_\phi$. This choice retains the global holonomy data along the non-contractible circle while reducing the bulk theory to an effectively two-dimensional BF theory.

\section{Constrained dimensional reduction}
\label{sec:reduction}
The dimensional reduction of the $AdS_3$ action \eqn{eq:gravascssec2} is implemented  by requiring the gauge connection $A$ ($\bar{A}$) to be invariant along the flow generated by the vector field $X=\partial_\phi$, that is
by imposing  vanishing Lie derivative :
\begin{equation}
\mathfrak L_X A=0 
\label{eq:LXconstraintsec3}
\end{equation}
that, in  adapted coordinates, reads
\begin{equation}
\partial_\phi A_\mu=0 .
\end{equation}
Let us notice that this condition could be relaxed by requiring instead
\begin{equation}
     \mathfrak{L}_XA = D_A \epsilon,
    \label{covariantred}
\end{equation}
where $\epsilon=\epsilon(\phi)$ is an infinitesimal gauge parameter. This would imply that the effect of an infinitesimal diffeomorphism along the reducible dimension could be reabsorbed by a local  transformation $h(\phi)=\exp[\epsilon(\phi)]$ of the gauge connection. Namely, the general $\phi$-dependent ${A}^h$ is just given by a gauge transformation on the $\phi$-independent connection $A$  which satisfies \eqn{eq:LXconstraintsec3}. We will keep working with the condition \eqn{eq:LXconstraintsec3} since the Chern-Simons action for ${A}^h$ only differs by  a boundary term
\be
  {S}[A^h] = S[A]+\int_{\partial M_3}\operatorname{Tr}(A\wedge h^{-1}dh).
\ee
 Assuming \eqn{eq:LXconstraintsec3} then corresponds to ignore the additional boundary degree of freedom $h$; this   amounts to choose  a boundary condition where $h$ is trivial on $\partial M_3$. Analogous conditions hold for $\bar A$.

As a further remark, let us  note that Eq. \eqref{eq:LXconstraintsec3} does \emph{not} imply that the holonomy of the connection 
\begin{equation}
\Hol_\phi(A)=P\exp\oint_{S^1} A_\phi\,d\phi
\label{eq:holphisec3}
\end{equation}
along the $\phi$-cycle is trivial.  Thus, the reduced sector retains global information associated with the non-contractible circle. 

Since the two  sectors are formally equivalent, from now on  we shall focus on a single $\mathfrak{sl}(2,\mathbb R)$ connection $A$, being understood that the same analysis applies to $\bar A$, unless otherwise stated.
\subsection{Symmetry constraint and constrained equations of motion}
It is convenient to implement the constraint \eqref{eq:LXconstraintsec3} dynamically. To this, we introduce a Lie-algebra-valued two-form  $\zeta$, that plays the role of a Lagrange multiplier, and we consider the constrained action
\begin{equation}
S[A,\zeta]
=
S_{\CS}[A]-\int_{M_3}\operatorname{Tr}\!\left(\zeta\wedge \mathfrak{L}_X A\right),
\label{eq:constrainedactionsec3}
\end{equation}
where $S_{\CS}[A]$ is the Chern--Simons functional \eqref{eq:csactionsec2}. Varying with respect to $A$ and $\zeta$, and using the standard variation of the Chern--Simons action, one finds
\begin{align}
\delta S
&=
\frac{k_{CS}}{2\pi}\int_{M_3}\operatorname{Tr}(\delta A\wedge F)
-\frac{k_{CS}}{4\pi}\int_{\partial M_3}\operatorname{Tr}(A\wedge \delta A)
\nonumber\\
&\qquad
-\int_{M_3}\operatorname{Tr}\!\left(\delta\zeta\wedge \mathfrak{L}_X A\right)
-\int_{M_3}\operatorname{Tr}\!\left(\zeta\wedge \mathfrak{L}_X\delta A\right).
\label{eq:variationconstrainedsec3a}
\end{align}
Using Cartan’s formula and integrating the last term by parts along the $X$-direction, one obtains
\begin{equation}
\delta S
=
\int_{M_3}\operatorname{Tr}\!\left[
\delta A\wedge \left(\frac{k}{2\pi}F+\mathfrak{L}_X\zeta\right)
-\delta\zeta\wedge \mathfrak{L}_X A
\right]
+\delta S_{\partial M_3},
\label{eq:variationconstrainedsec3b}
\end{equation}
where $\delta S_{\partial M_3}$ collects the boundary contributions. The corresponding bulk equations of motion are 
\begin{equation}
\mathfrak{L}_X A=0,
\qquad
\frac{k_{CS}}{2\pi}F+\mathfrak{L}_X\zeta=0.
\label{eq:constrainedeomsec3}
\end{equation}
Thus the Lagrange multiplier enforces the symmetry constraint, while the curvature is determined by the Lie derivative of  $\zeta$ with respect to $X$.
If one further imposes
\begin{equation}
\mathfrak{L}_X\zeta=0 ,
\label{eq:LXlambdasec3}
\end{equation}
then the second equation in \eqref{eq:constrainedeomsec3} gives back  the flatness condition
\begin{equation}
F=0.
\label{eq:flatnesssubsecorsec3}
\end{equation}
This setting is the one relevant for the induced boundary dynamics analyzed later.

\subsection{Geometric decomposition adapted to the reduction}

To perform the reduction explicitly, it is convenient to decompose the connection according to the vector field $X$.  Let $X^\flat$ denote a one-form dual to $X$, normalized by
\begin{equation}
\iota_X X^\flat =1,
\qquad
\iota_X dX^\flat=0 .
\label{eq:Xflatnormsec3}
\end{equation}
With the product structure $M_3=M_2\times S^1$ and the adapted coordinates \eqn{coordinates} we have
\begin{equation}
X^\flat=d\phi,
\qquad
dX^\flat=0 .
\label{eq:Xflatdphisec3}
\end{equation}
The connection can then be decomposed as
\begin{equation}
A=\Phi\,X^\flat+A_\perp,
\qquad
\Phi=\iota_X A,
\qquad
\iota_X A_\perp=0 
\label{eq:Adecompsec3}
\end{equation}
that is, 
\begin{equation}
A=\Phi\,d\phi+A_\rho\,d\rho+A_\tau\,d\tau 
\label{eq:Adecompcoordsec3}
\end{equation}
with $
A_\perp=A_\rho\,d\rho+A_\tau\,d\tau $. 
Thus $\Phi$ is a Lie-algebra-valued scalar on $M_2$, while $A_\perp$ is a connection one-form on the reduced two-dimensional space $M_2$. More generally, given the field $X$, any differential form $B$ can be written as
\begin{equation}
B=B_\perp+X^\flat\wedge B_X,
\qquad
B_X=\iota_X B,
\qquad
\iota_X B_\perp=0 .
\label{eq:Bdecompsec3}
\end{equation}
The exterior derivative correspondingly splits as
\begin{equation}
d=d_\perp+X^\flat\wedge \mathfrak{L}_X ,
\label{eq:dsplitsec3}
\end{equation}
where $d_\perp$ acts only along the directions transverse to $X$. Since $X^\flat=d\phi$ is closed, no additional curvature terms appear in this decomposition.
Under the symmetry condition $\mathfrak{L}_X A=0$, both $A_\perp$ and $\Phi$ are independent of $\phi$. This is immediate from \eqref{eq:Adecompsec3}, since
\begin{equation}
\mathfrak{L}_X A
=
(\mathfrak{L}_X \Phi)\,d\phi+\mathfrak{L}_X A_\perp ,
\end{equation}
and therefore
\begin{equation}
\mathfrak{L}_X A=0
\qquad\Longrightarrow\qquad
\mathfrak{L}_X\Phi=0,
\qquad
\mathfrak{L}_X A_\perp=0 .
\label{eq:LXcomponentssec3}
\end{equation}

\subsection{Reduced curvature and emergence of the BF-like system}

We now compute the curvature in terms of the decomposed fields. Starting from
\begin{equation}
A=\Phi\,d\phi+A_\perp ,
\end{equation}
one has
\begin{align}
dA
&=
d_\perp\Phi\wedge d\phi+d_\perp A_\perp+d\phi\wedge \mathfrak{L}_X A_\perp .
\label{eq:dAsec3}
\end{align}
Moreover,
\begin{align}
A\wedge A
&=
(\Phi\,d\phi+A_\perp)\wedge(\Phi\,d\phi+A_\perp)
\nonumber\\
&=
A_\perp\wedge A_\perp
+\Phi\,d\phi\wedge A_\perp
+A_\perp\wedge \Phi\,d\phi 
\end{align}
where  the mixed terms combine into
\begin{equation}
\Phi\,d\phi\wedge A_\perp+A_\perp\wedge \Phi\,d\phi
=
-d\phi\wedge [A_\perp,\Phi]
\label{eq:mixedtermssec3}
\end{equation}
so that 
\begin{equation}
A\wedge A
=
A_\perp\wedge A_\perp-d\phi\wedge [A_\perp,\Phi].
\label{eq:AAsec3}
\end{equation}
Combining \eqref{eq:dAsec3} and \eqref{eq:AAsec3}, one finds
\begin{align}
F
&=
dA+A\wedge A
\nonumber\\
&=
d_\perp A_\perp+A_\perp\wedge A_\perp
+d_\perp\Phi\wedge d\phi
+d\phi\wedge \mathfrak{L}_X A_\perp
-d\phi\wedge [A_\perp,\Phi]\nonumber\\
&=
F_\perp-d\phi\wedge \bigl(d_\perp\Phi+[A_\perp,\Phi]-\mathfrak{L}_X A_\perp\bigr),
\label{eq:Fbeforeconstraintsec3}
\end{align}
where
\begin{equation}
F_\perp=d_\perp A_\perp+A_\perp\wedge A_\perp .
\label{eq:Fperpdefsec3}
\end{equation}
Once the symmetry condition $\mathfrak{L}_XA_\perp=0$ is imposed, this simplifies to
\begin{equation}
F
=
F_\perp-d\phi\wedge D_\perp\Phi ,
\label{eq:Fdecompfinalsec3}
\end{equation}
with
\begin{equation}
D_\perp\Phi=d_\perp\Phi+[A_\perp,\Phi].
\label{eq:DperpPhisec3}
\end{equation}
Equation \eqref{eq:Fdecompfinalsec3} shows that, after imposing the symmetry, the three-dimensional 
curvature factorizes, with $F_\perp$ the curvature of the 2-dimensional connection, $A_\perp$, on $M_2$, and $D_\perp\Phi$ the covariant derivative of the  adjoint scalar $\Phi$.  In particular, the flatness condition $F=0$ is equivalent to a pair of two-dimensional equations
\begin{equation}
F_\perp=0,
\qquad
D_\perp\Phi=0 .
\label{eq:2dBFeomssec3}
\end{equation}
These are precisely the equations of motion of a BF  model in two dimensions, \cite{Isler:1989hq,Cangemi:1992bj,Ikeda:1993fh,Schaller:1994es}, with $\Phi$ playing the role of the $B$ field.
In the next section, we analyse the  variational principle of the three dimensional model with  admissible boundary data.  

\section{Dimensional reduction of the action and the pure-gauge sector}
\label{sec:actionreduction}
The geometric decomposition of the connection and curvature derived in Section~\ref{sec:reduction} can now be inserted directly into the Chern--Simons action. This yields the exact reduced BF-like action on $M_2$ plus boundary term. Indeed, using 
\begin{equation}
A=A_\perp+\Phi\,d\phi,
\qquad
\Phi=\iota_{\partial_\phi}A,
\qquad
\iota_{\partial_\phi}A_\perp=0,
\label{eq:Adecompsec5exact}
\end{equation}
and imposing the symmetry condition
$\mathfrak L_{X}A=0, $
one finds
\begin{equation}
A\wedge dA+\frac23 A\wedge A\wedge A
=
d\phi\wedge
\operatorname{Tr}\!\left(
2\Phi F_\perp-d_\perp(\Phi A_\perp)
\right)
\label{eq:CSdecompidentity}
\end{equation}
then, integrating over $\phi$
%
we obtain 
\begin{align}
S_{\CS}[A]
&=
\frac{k_{CS}}{4\pi}\int_{S^1}d\phi\int_{M_2}
\operatorname{Tr}\!\Bigl(
2\Phi F_\perp-d_\perp(\Phi A_\perp)
\Bigr)
\nonumber\\
&=
k_{CS}\int_{M_2}\operatorname{Tr}(\Phi F_\perp)
-\frac{k_{CS}}{2}\int_{\partial M_2}\operatorname{Tr}(\Phi A_\tau)\,d\tau 
\label{eq:Sredexact}
\end{align}
that is the reduced action, with normalization inherited directly from the original Chern--Simons prefactor.

\subsection{Flatness condition and pure-gauge parametrization}
We now restrict to the flat solution $F=0$ of the 3d action 
and perform the reduction.
Locally, the flatness condition implies that the full connection is pure gauge:
\begin{equation}
A=g^{-1}dg,
\qquad
g\in SL(2,\mathbb R)
\label{eq:puregaugesec5exact}
\end{equation}
 namely, at the boundary
\begin{equation}
\Phi=g^{-1}\partial_\phi g,
\qquad
A_\tau=g^{-1}\partial_\tau g \label{eq:PhiAtauexact}
\end{equation}
while the symmetry condition $\mathfrak{L}_{X}A=0$ implies
\begin{equation}
\partial_\phi(g^{-1}\partial_\mu g)=0, \qquad \mu=\phi, \tau
\label{eq:phiinvarianceg}
\end{equation}
and we recall that analogous results hold for   $\bar A$.

\subsection{One-dimensional boundary action}
On using Eqs. \eqn{eq:2dBFeomssec3} the bulk BF term in the reduced action \eqref{eq:Sredexact} vanishes identically, therefore we obtain
\begin{equation}
S_{CS}\big|_{F_\perp=0}
=
-\frac{k_{CS}}{2}\int_{\partial M_2}\operatorname{Tr}(\Phi A_\tau)\,d\tau 
\label{eq:Sredflatexact}
\end{equation}
and Eqs.  \eqref{eq:PhiAtauexact} yield
\begin{equation}
S_{CS}\big|_{F_\perp=0}
=
-\frac{k_{CS}}{2}\int_{S^1}
\operatorname{Tr}\!\left(
g^{-1}\partial_\phi g\, g^{-1}\partial_\tau g
\right)d\tau .
\label{eq:Sredflatgexact}
\end{equation}
The same result is obtained directly from the Chern--Simons action evaluated on-shell, with $A=g^{-1}dg$. Indeed, for a pure gauge connection one has
\begin{equation}
S_{\CS}[g]
=
-\frac{k_{CS}}{12\pi}\int_{M_3}\operatorname{Tr}\!\left((g^{-1}dg)^3\right).
\label{eq:SCSpuregaugeexact}
\end{equation}
Using the decomposition $g^{-1}dg=A_\perp+\Phi\,d\phi$ together with $\partial_\phi(g^{-1}\partial_\mu g)=0$, the integrand reduces to
\begin{equation}
\operatorname{Tr}\!\left((g^{-1}dg)^3\right)
=
3\,d\phi\wedge \operatorname{Tr}\!\left(\Phi\,A_\perp\wedge A_\perp\right).
\end{equation}
Since $F_\perp=0$, we have
\begin{equation}
A_\perp\wedge A_\perp=-d_\perp A_\perp,
\end{equation}
and therefore
\begin{equation}
\label{eq:CSdirectexact}
S_{\CS}[g]
=
-\frac{k_{CS}}{12\pi}\int_{M_3}3\,d\phi\wedge \operatorname{Tr}\!\left(\Phi\,A_\perp\wedge A_\perp\right)
=-\frac{k_{CS}}{2}\int_{\partial M_2}\operatorname{Tr}(\Phi A_\tau)\,d\tau 
\end{equation}
which agrees with Eq. \eqref{eq:Sredflatexact}. Therefore the latter  represents the  one-dimensional boundary action inherited from the symmetry-reduced Chern--Simons theory. 
The emergence of one-dimensional geometric actions from Chern--Simons and BF-type theories is consistent with previous work in low-dimensional gravity \cite{Blommaert:2018oro,Valach:2019arb,Barnich:2017jgw}.
The reduced boundary theory acquires the form \eqn{eq:Sredflatexact} without further assumptions. 

\section{Variational Principle}
\label{sec:vp}
In the previous sections we showed that, after imposing the symmetry constraint along the non-contractible circle and integrating over the $\phi$ direction, the three-dimensional Chern--Simons theory reduces to a two-dimensional BF model on $M_2$. Moreover, on restricting to  flat connections dictated by the  {on-shell condition}, 
one obtains  a  one-dimensional boundary action.
Let us analyze the corresponding variational principle and focus on possible extra boundary contributions. 
As previously,
we consider  a single $\mathfrak{sl}(2,\mathbb R)$ connection, being understood that the same procedure and results hold for the other $SL(2,\R)$ sector  of the action unless otherwise explicitly stated.  
The variation of the Chern--Simons action \eqref{eq:csactionsec2}
reads
\begin{equation}
\delta S_{\CS}[A]=\frac{k_{CS}}{2\pi}\int_{M_3}\operatorname{Tr}(\delta A\wedge F)-\frac{k_{CS}}{4\pi}\int_{\partial M_3}\operatorname{Tr}(A\wedge \delta A).
\end{equation}
 This reduces on shell to the boundary contribution
\begin{equation}
\delta S_{\CS}[A]\big|_{\mathrm{on\ shell}}=-\frac{k_{CS}}{4\pi}\int_{\partial M_3}\operatorname{Tr}(A\wedge \delta A).
\end{equation}
Using the boundary restriction of the decomposition in \eqref{eq:Adecompsec5exact}, 
 the boundary contribution of the variation reads 
\begin{equation}
    \delta S_{CS}[A]_{\text{on shell}} =- \frac{k_{CS}}{4\pi}\int_{\partial M_3} \operatorname{Tr}(A_{\tau}\delta\Phi - \Phi \delta A_{\tau})\, d\tau  d\phi. \label{eq:deltaSredonshell}
\end{equation}
In order to have a well posed variational principle, we need to specify boundary conditions for $A_{\tau}$ and $\Phi$ and possibly introduce additional boundary terms for the Chern--Simons action. 
A natural choice is to restrict the boundary fields to the subspace
\begin{equation}
{(A_\tau)}_{\partial M_3 }=\lambda'(\tau)\Phi
\label{noncovariant}
\end{equation}
with $\lambda$ an orientation preserving diffeomorphism of $\partial M_2$.
Hence, the admissible variations satisfy $\delta A_\tau=\lambda'\delta\Phi+ \delta\lambda' \Phi.$ This is   the standard choice for $AdS_3$ gravity. Indeed, imposing similar boundary conditions on the other  sector leads to two copies of  the 2d Wess-Zumino-Witten action, which in turn reduce to the Liouville theory on the two-dimensional boundary \cite{Coussaert:1995zp}. 
The boundary condition \eqn{noncovariant} is not gauge covariant since, after dimensional reduction, the $\Phi$ field ceases to transform as a gauge connection, transforming instead in the adjoint representation. A more general family of boundary conditions is of the form
\begin{equation}
{(A_\tau)}_{\partial M_3 }=\lambda'\Phi+ u^{-1}\partial_\tau u,
\label{eq:BC2new}
\end{equation}
 with $u: S_1\rightarrow SL(2,\mathbb R)$  a  group-valued field, similar in spirit to the introduction of the edge modes  in \cite{Valach:2019arb} (see also \cite{Attard:2017sdn}). 
In order for the variation in \eqref{eq:deltaSredonshell} to be set to zero by the covariant conditions \eqn{eq:BC2new}, a natural prescription for the boundary variations is 
\begin{equation}
    \delta\Phi =0.
\end{equation}
Therefore, the only degrees of freedom are represented by the time-circle reparametrisation $\lambda$ and the  edge modes $u$. Compatibility of \eqref{eq:BC2new} with the equation of motion \eqref{eq:2dBFeomssec3} imposes 
\begin{equation}
\label{Phisol}
    \Phi = u^{-1}Tu
\end{equation}
with $T$ a Lie algebra element as shown in the Appendix~\ref{app:covBC}.  This implies that the  connection $A_{\tau}$ is given 
at the 2D boundary by a gauge transformation $u$ acting on a reference configuration 
\be\label{referenceA}
A_{\tau}= u^{-1}(A^{0}_{\tau}+\partial_{\tau})u, \quad
A_{\tau}^0 = \lambda' T.
\ee
Therefore, the variational problem is not formulated on the full space of independent boundary values of $(A_\tau,\Phi)$, but on the restricted  subspace identified by Eq. \eqref{eq:BC2new}. Equivalently, one can regard \eqref{eq:BC2new} as specifying the admissible boundary configurations, parametrized by the pair $(\lambda,u)$, on which the Chern-Simons action is differentiable. 

In two dimensions, covariant and non covariant conditions lead to the same counter-term needed to absorb the spurious boundary variation of the bulk BF theory. In three dimensions, although leading to different action functionals at the boundary, they are expected to be equivalent up to reparametrisation. In the remainder of the paper we will stick to the boundary dynamics associated with the non-covariant prescription.  
\subsection{3D Boundary conditions}
So far, we have just introduced two different ways to ensure a well posed variational principle, but further boundary conditions have to be imposed in order   to characterize the three-dimensional geometry we want to reduce. A common set of boundary conditions in three-dimensional gravity are the Brown-Henneaux  ones \cite{Brown:1986nw, Coussaert:1995zp}, where in a given gauge
\begin{align}
    &A^{(+)} = -\bar{A}^{(-)}= \text{const.},\quad A^{(0)} = \bar{{A}}^{(0)}=0, \nonumber \\ &A^{(-)}, \bar{A}^{(+)} = \text{arbitrary functions of $(\tau, \phi)$}, 
    \label{bhcond}
\end{align}
and the connection $A $ (resp. $\bar A$) is represented in   the standard $\mathfrak{sl}(2,\mathbb{R})$  basis,  $\big\{L_{\pm}, L_0 \big\}, $    specified in \eqn{eq:sl2basissec6}. Those boundary conditions are derived by requiring an asymptotically $AdS_3$ spacetime in the metric formulation. As shown in \cite{Donnay:2016iyk}, they can be realized in a given frame where, given $x^{\pm}=\tau\pm \phi$, one gets
\begin{equation}
    A_{x^{+}} = L_{+} + \mathcal{L}(x^{+})L_{-}, \quad \bar{
    A}_{x^{-}} = \bar{\mathcal{L}}(x^{-})L_{+} + L_{-}.
    \label{DSform2}
\end{equation}
where $\mathcal{L}, \bar{\mathcal{L}}$ are arbitrary functions on $\partial M_3$.  It is important to notice that the fall-off conditions leading to Eqs. \eqn{bhcond} do not fix the parametrization of the boundary, therefore, if Eqs. \eqn{bhcond} hold in some frame, a naive reparameterization $x^{\pm} \rightarrow f^{\pm}(x^{\pm})$ of the boundary would not preserve the prescriptions \eqn{DSform2}. 
 Hence, in order for  \eqn{bhcond} to be properly  formulated, we need to require that any reparameterization can be reabsorbed by a suitable gauge transformation. Given for example $\tilde{x}^{+}=f(x^{+})$, one has
 \begin{eqnarray}
     A_{\tilde{x}^{+}} = {f'(x^{+})}A_{x^{+}},
 \end{eqnarray}
which no longer satisfies the boundary prescriptions. Nevertheless, the effect of the reparametrisation can be reabsorbed by a gauge transformation by requiring
\begin{eqnarray}
    A^u_{\tilde{x}^{+}} = u^{-1}(A_{\tilde{x}^{+}}  + \partial_{\tilde{x}})u = L_{+} + \widetilde{\mathcal{L}}(\tilde{x}^{+})L_{-},
\end{eqnarray}
 which implies that $\mathcal{L}(x^{+})$ must transform as a dual element of the Virasoro algebra under the action of a diffeomorphism
 \begin{eqnarray}
     \widetilde{\mathcal{L}}(f({x}^{+})) ={\mathcal{L}}(x^{+})f'(x^{+})^2 + \frac{1}{2} \{f(x^{+}),x^{+}\}_S.
 \end{eqnarray}
This is why these boundary conditions correspond to an asymptotic symmetry algebra given by two independent copies of the Virasoro algebra. 
The relation between reparameterizations and gauge transformations needed to restore the \eqn{DSform2} simplifies once the gauge connection is considered on-shell. In fact, given $\xi$ as the infinitesimal generator of a reparameterization, the following identity holds
\begin{equation}
    \mathfrak{L}_{\xi}A = D_{A}(\iota_\xi A) - i_\xi F,
\end{equation}
which on-shell reduces to
$\mathfrak{L}_{\xi}A = D_{A}(\iota_\xi A).
$ Diffeomorphisms can then be recast as gauge transformations generated by gauge parameters $\epsilon= \iota_\xi A$. In this sense, once on-shell, searching for the gauge transformations that restore \eqn{DSform2}, when a given reparameterization is performed, is equivalent to just ask for the general gauge transformation preserving it. 

We will be interested in a different asymptotic symmetry algebra, given by the semidirect sum of a Virasoro algebra with  a $\mathfrak{sl(2,\mathbb{R}})$-Kac-Moody  algebra, as in \cite{Chirco:2024ubu, Gonzalez:2018enk}.  Our proposal  represents  a further extension of the algebra of asymptotic symmetries of \cite{Compere:2013bya} where the Kac-Moody sector is Abelian. We consider this requirement  necessary to map three-dimensional gravity to generalized SYK models as for example in \cite{Yoon:2017nig} and, therefore, we are  led to introduce a different set of boundary conditions. For a complete compendium of the most general boundary conditions in $AdS_3$ generalizing the Brown-Henneaux ones, see e.g.  \cite{Grumiller:2016pqb}.  

The boundary conditions we are interested in correspond to 
\begin{equation}
\label{drinfeldsokolov}
    A^{(+)} = \text{const.},\quad A^{(0)} = 0, \quad A^{(-)} = \mathcal{L}(\tau, \phi),
\end{equation}
together with 
\begin{equation}
\label{dressing}
    \bar{A}= g^{-1}dg, \quad g(\sigma, \phi) = w(\sigma, \phi)b(\tau(\sigma)),
\end{equation}
where $\sigma= \lambda(\tau)$, the reparametrization mode  in Eqs. \eqn{noncovariant},  \eqn{eq:BC2new}. This choice corresponds to imposing Brown-Henneaux conditions on a single $\mathfrak{sl}(2,\mathbb{R})$ sector while keeping the other free and they  yield  the same boundary dynamics provided by the conditions  (4.23) and (4.24) of \cite{Gonzalez:2018enk}, which in turn are the first order conditions of  \cite{Avery_2014}. The dressing mode $b$ is necessary in order to couple the two sectors and get the semidirect product structure. As we will show, the dimensional reduction of Chern-Simons theory equipped with these prescriptions will lead to a  boundary dynamics candidate to be dual to SYK tensor models with $\SL(2,\R)$ global symmetry similarly to what shown in \cite{Yoon:2017nig}.

\section{Drinfel'd--Sokolov reduction}
\label{sec:deformed}
Let us consider the boundary conditions introduced  in  section~\ref{sec:vp} namely Eq. \eqn{noncovariant}. 
We are interested in the reduction of the one-dimensional action \eqref{eq:Sredflatexact}, once Eqs. \eqn{drinfeldsokolov} are imposed for a single $\mathfrak{sl}(2,\mathbb{R})$ sector and dimensionally reduced.
This is the starting point for the Drinfel'd--Sokolov construction, corresponding to an asymptotic $AdS_2$ behavior of the geometry near $\partial M_2$ \cite{Valach:2019arb,Blommaert:2018oro}.
We use the standard highest-weight basis

\begin{equation}
L_0=\frac12
\begin{pmatrix}
1 & \,\,\,\,0\\
0 & -1
\end{pmatrix},
\qquad
L_-=
\begin{pmatrix}
0 & -1\\
0 & \,\,\,\,0
\end{pmatrix},
\qquad
L_+=
\begin{pmatrix}
0 & \,\,\,\,0\\
1 & \,\,\,\,    0
\end{pmatrix},
\label{eq:sl2basissec6}
\end{equation}
satisfying
\begin{equation}
[L_0,L_\pm]=\mp L_\pm,
\qquad
[L_+,L_-]=2L_0,
\end{equation}
with invariant bilinear form given by the trace in the fundamental representation,
\begin{equation}
\operatorname{Tr}(L_0^2)=\frac12,
\qquad
\operatorname{Tr}(L_+L_-)=-1,
\qquad
\operatorname{Tr}(L_\pm^2)=0.
\label{eq:traceconvsec6}
\end{equation}
 It turns out that a residual gauge symmetry is recovered when the boundary connection is constrained to the configuration space defined by \eqn{drinfeldsokolov}, given by local transformations generated by $L_{-}$ \cite{Valach:2019jzv}. In our case, as a consequence of the dimensional reduction in Section~\ref{sec:actionreduction}, we consider a local version of the boundary condition
 \eqn{drinfeldsokolov} up to some gauge transformation $u$,
\begin{eqnarray}
    A^{u}_{\tau}= L_{+} + \mathcal{L}(\tau)L_{-},
    \label{DSform}
\end{eqnarray}
where the dependence on $\phi$ has been removed. 
We will refer to the latter as Drinflel'd-Sokolov gauge. Starting from \eqref{noncovariant} and given the equation of motion $D_A\Phi=0$, we have 
\begin{eqnarray}
    \Phi'=0.
\end{eqnarray}
This means that $A_\tau$ is in the DS form just up to a reparameterization, provided that 
\be
\label{phieq}
\Phi = L_{+}-\frac{C}{2}L_{-}.
\ee
In order to bring it back in the DS form, we need to introduce a gauge transformation such that
\begin{equation}
    A^u_{\tau}= u^{-1}(A_{\tau}+ \partial_{\tau})u
    \label{gaugeDS}
\end{equation}
is in the form \eqn{DSform}. This is achieved by 
\begin{eqnarray}
\label{eq:u}
    u = \begin{pmatrix}
        (\lambda')^{-\tfrac{1}{2}} & -\tfrac{1}{2}(\lambda')^{-\tfrac{3}{2}}\lambda'' \\ 0 & (\lambda')^{\tfrac{1}{2}}
    \end{pmatrix},
\end{eqnarray}
while for $\mathcal{L}(\tau)$ we find
\begin{eqnarray}
\label{Ftau}
    \mathcal{L}(\tau) = -\frac{C}{2}\lambda'^2 +\frac{1}{2} \{ \lambda, \tau\}_S,
\end{eqnarray}
that is, in a generic frame, $\mathcal{L}(\tau)$ can be identified with the result of the coadjoint action of the finite diffeomorphism  $\lambda$ on  $-C/2$, a constant element in the dual of the Virasoro algebra, labelling    coadjoint orbits. The choice of $C$ sets the particular orbit. 
Summarizing, the reduction of the action  under the  boundary condition \eqref{noncovariant} leads to
\begin{equation}\label{redac}
    S_{1D}[\lambda]=-\frac{k_{CS}}{2}\int_{S^1}\lambda'\operatorname{Tr}(\Phi)^2d\tau, \end{equation}
which, by virtue of Eqs. \eqn{phieq}, \eqn{Ftau}, can be rewrtitten as
\begin{eqnarray}\label{rewr}
    S_{1D}[\lambda]= k_{CS}\int_{S^1}\Big( \frac{\mathcal{L}(\tau)}{\lambda'} -\frac{1}{2\lambda'}\{\lambda,\tau\}_S \Big)d\tau.
\end{eqnarray}
In order to compare with standard notation, it is convenient to reparametrize $\tau=\lambda^{-1}(\sigma)$, so to have
\be\label{change}
\lambda'(\tau)= \frac{1}{\tau'(\sigma)}, \;\;\;\; \{\lambda, \tau\}_S= -\frac{1}{(\tau'(\sigma))^2}\{\tau,\sigma\}_S,\;\;\; d\tau= \tau'(\sigma) d\sigma
\ee
and 
\begin{eqnarray}\label{standard}
    S_{1D}[\lambda]= k_{CS}\int_{S^1}\Big(\tau'(\sigma)^2 \mathcal{L}(\tau) + \frac{1}{2}\{\tau, \sigma \}_S \Big)d\sigma.
\end{eqnarray} 
By virtue of the composition rule of the Schwarzian derivative,   the latter can be rewritten in terms of a single Schwarzian derivative only when   $\mathcal{L}(\tau)=\frac{1}{2}\{F(\tau),\tau\}_S$. We show in Appendix \ref{appendixA} that  this is the case provided $A^u_{\tau}$ is pure gauge. 
Then,  the action \eqn{standard} acquires the form
\begin{eqnarray}
\label{Schwarziandef}
    S_{1D}[\lambda]&=& k_{CS}\int_{S^1}\Big(\tau'(\sigma)^2\frac{1}{2}\{F(\tau), \tau \}_S + \frac{1}{2}\{\tau, \sigma \}_S \Big)d\sigma \nonumber \\&=& \frac{k_{CS}}{2}\int_{S^1}\{F\circ\tau(\sigma), \sigma \}_S \,d\sigma.
\end{eqnarray}
As we shall see, for exceptional orbits labeled by the value of the Casimir $C$,  one finds explicitly  $F=\tan( \sqrt{|C|/2}\lambda)$, up to global SL$(2,\R)$ transformations. The value of $C$, in turn,  is selected by holonomy conditions. Let us see how.
  
We recall that $C$ is a constant of motion on-shell, it being the Casimir function $C=\operatorname{Tr}(\Phi)^2$. Compatibly with our choice of the 3D manifold as $AdS_3$, we require (see e.g. \cite{Gonzalez:2018enk}) that the holonomy $\operatorname{Hol}_{\tau}(A_{\tau})$ be trivial along the contractible circle
\begin{equation}
\label{holcond}
    \operatorname{Hol}_{\tau}(A_{\tau}) := \mathcal{P}\exp \left( \int_{S^1}A_{\tau}d\tau \right)= \pm\mathbb{I}.
\end{equation}
Since $\lambda$  must be a well defined diffeomorphism of the circle, it has to satisfy
\begin{eqnarray}
    \lambda(2\pi)-\lambda(0)=2\pi,
\end{eqnarray}
therefore
\begin{eqnarray}
\label{phiconst}
    \operatorname{Hol}_{\tau}(A_{\tau}) = e^{2\pi \Phi}=\pm \mathbb{I}, \quad \text{with} \quad \Phi = L_{+}-\frac{C}{2}L_{-}.
\end{eqnarray}
 Then, the explicit form of $\Phi$ in the basis \eqn{eq:sl2basissec6} is 
\begin{equation}
    \Phi= \begin{pmatrix}
      \,  0\, &\,\, C/2\\ \,1&0
    \end{pmatrix}.
\end{equation}
Let $\Phi_d = \text{diag}(\omega,-\omega)$ be the diagonalized $\Phi$ matrix,  we have
\begin{eqnarray}\label{integern}
    2\pi\,\omega = in\pi, \quad n\in\mathbb{Z}, n\neq0,
\end{eqnarray}
 which implies
\begin{eqnarray}
\label{eq:fixC}
    C = -n^2/2, \;\;\; n\neq 0.
\end{eqnarray}
Therefore, the holonomy condition leads to  \begin{eqnarray}
\label{coadjointactionS}
    \mathcal{L}(\tau) = \frac{n^2}{4}\lambda'^2 + \frac{1}{2}\{\lambda, \tau\}_S 
\end{eqnarray}
which can be rewritten as a Schwarzian derivative 
\begin{equation}
    \mathcal{L}(\tau) = \frac{1}{2}\{F(\lambda(\tau)), \tau\}_S,
\end{equation}
with 
\be
F= \tan(\frac{n}{2}\lambda(\tau)),
\ee
up to global SL$(2,\R)$ transformations.
Eq. 
\eqn{eq:fixC} for the value of the Casimir function selects the exceptional orbits \cite{Witten:1987av}:
\begin{eqnarray}
    \operatorname{Diff}(S^1)/\operatorname{SL}^{(n)}(2,\mathbb{R}).
\end{eqnarray}

If we replace the unit $2\pi$-circle by a physical thermal circle with periodicity $\tau+\beta \sim \tau$, we get
\begin{eqnarray}
    C = -\frac{2\pi^2n^2}{\beta^2},
\end{eqnarray}
together with the overall rescaling of $\mathcal{L}(\tau)$ of the same factor $\frac{4\pi^2}{\beta^2}$. This corresponds to the thermal exceptional orbits relevant for the Schwarzian sector of JT gravity \cite{Mertens:2022mtm}. Notice that the case $n=0$, forbidden by Eq. \eqn{integern}, namely violating the holonomy condition \eqn{holcond}, would  correspond to $C=0$, namely, $\Phi\propto L_+$, that is the  zero temperature degenerate limit, $\beta\rightarrow\infty$. The related orbit is in this case $\operatorname{Diff}(S^1)/\operatorname{S}^{1}$ \cite{Witten:1987av}.

\section{Current extension of the Schwarzian theory}
\label{sec:KMextensions}
In the previous two sections we derived a one dimensional boundary theory on the coadjoint orbits of the Virasoro group, among which the exceptional orbits giving rise to JT gravity, from the symmetry-reduced $AdS_3$ Chern-Simons action. We showed how this reduction is consistent with dimensionally reduced Brown-Henneaux boundary conditions for a single $\mathfrak{sl}(2, \mathbb{R})$ sector. 

In this section we address a second  question, namely the possibility of allowing  current degrees of freedom at the boundary. As anticipated in the introduction, this is an interesting problem, related to the emergence of additional degrees of freedom in the  so-called tensor SYK models \cite{Yoon:2017nig}, especially within their low energy limit where they are expected to be described by a Schwarzian action plus Kac-Moody currents contributions. These models  describe the dynamics of  $N$ Majorana fermions $\psi_i^a$, with random, all-to-all four-body interactions, which besides the index $i$ running from $1$ to $N$, carry  a vector index, $a$,   with respect to some Lie algebra. From the gravitational $AdS_3$ reduction analyzed  so far, it is then natural to investigate whether the new degrees of freedom can be traced back to the same parent action in 3d. 
 As we will see, the currents extension is controlled by how the second  $\SL(2,\mathbb R)$ sector for the connection $\bar A$  is organized after the reduction. 
At the level of the bulk gauge algebra one has the standard decomposition
\begin{equation}
\mathfrak{so}(2,2)\simeq \mathfrak{sl}(2,\mathbb R)\oplus \mathfrak{sl}(2,\mathbb R).
\label{eq:so22directsumsec8}
\end{equation}
If the two sectors are treated  as independent copies, then after reduction one obtains two decoupled one-dimensional models. In that case the second sector contributes only an additional set of boundary degrees of freedom, but does not couple to the first one. A different and more interesting possibility is to reorganize the second sector, after reduction, as a  loop-valued  sector, acted upon by the reparametrization mode of the first. In this effective boundary interpretation, one is naturally led to a semidirect-product structure between the reparametrization group of the first sector and a loop algebra based on $\SL(2,\mathbb R)$ \cite{Chirco:2024ubu}.
The purpose of this section is to describe this structure.  We will see how the Schwarzian sector, when suitably coupled with the other one, gives rise to the current-dressed Schwarzian theory in \cite{Chirco:2024ubu, Gonzalez:2018enk}.
\subsection{Schwarzian sector with current dressing}
We first consider the  gauge connection $A$, satisfying the boundary conditions \eqn{noncovariant}. As shown in Section~\ref{sec:deformed}, substituting this in the one-dimensional effective action \eqn{eq:Sredflatexact}
yields Eq. \eqn{change},
once \eqref{drinfeldsokolov} are imposed and dimensionally reduced.
Then, 
if the connection  $\bar A$,  is left independent, with the same kind of boundary condition, one simply obtains a second decoupled one-dimensional system. At the effective boundary level, however, one may instead regard the second sector as describing  an internal current degree of freedom transforming under the reparametrization mode $\lambda$ of the first sector \cite{Yoon:2017nig,Gross:2016kjj,Liu:2019niv,Narayan:2023wlk,Carrozza:2018psc,Gonzalez:2018enk,Chirco:2024so22}. Substituting $\bar A$ with the first of Eqs.  \eqref{dressing}, with $g:S^1\rightarrow \SL(2,\mathbb R)$, the corresponding group-valued boundary field, the  total dimensionally reduced boundary action acquires the form
\begin{equation}
S[\lambda,g]
=
\frac{k_{CS}}{2}\int_{S^1}\big( \mathcal{L}\tau'^2+\frac{1}{2}\{\tau,\sigma\}\big)  \,d\sigma
+ \frac{k_{CS}}{2}\int_{S^1}\operatorname{Tr}\,\bigl((g^{-1}g')^2\bigr)\,d\sigma 
\label{eq:uncoupledSKMsec8}
\end{equation}
where the boundary condition for the connection $\bar A$ has been chosen as
\begin{equation}
    \bar{A}_{\sigma}|_{S^1} = -\bar{\Phi}.
\end{equation}
It is important to notice that, by choosing $\bar{A}_{\sigma} = -\alpha\,\bar{\Phi}$ for some constant $\alpha$, the level $k_{CS}/2$ of the Kac-Moody sector can be tuned independently of the level $k_{CS}/2$ of the Schwarzian sector. A genuine coupling is obtained when the second sector is dressed by the reparametrization mode of the first. 
As anticipated by the second of Eqs. \eqref{dressing}, a  convenient way to encode this is through a field redefinition
\begin{equation}
g(\sigma)\longrightarrow w(\sigma)\,b(\tau(\sigma)),
\label{eq:dressingksec8}
\end{equation}
where $b(\tau(\sigma))$ is a fixed $\SL(2,\mathbb R)$-valued function of the Schwarzian mode. This expresses the fact that the second sector is no longer treated as  independent, as  the loop-valued field  is now acted upon by the reparameterization symmetry of the first sector. Under this dressing, the associated boundary  potential becomes
\begin{equation}
\bar{A}_\sigma
=
b^{-1}(w^{-1}w')b+b^{-1}\dot b\,\tau' ,
\label{eq:Atildesec8}
\end{equation}
where the dot denotes the derivative with respect to the argument  $\tau$. The resulting reduced action takes the form
\begin{equation}
S_{\mathrm{Sch+KM}}[\tau,w]
=
\frac{k_{CS}}{2}\int_{S^1}
\left[ \mathcal{L}\tau'^2+ \frac{1}{2}
\{\tau,\sigma\}
+\operatorname{Tr}\!\bigl((w^{-1}w')^2\bigr)
+2\tau'\operatorname{Tr}(P\,w^{-1}w')
+\tau'^2\,u(\tau)
\right]d\sigma ,
\label{eq:SchKMnormalizedsec8}
\end{equation}
where
\begin{equation}
P(\tau)=\dot b\,b^{-1},
\qquad
u(\tau)=\operatorname{Tr}(P^2).
\end{equation}
Up to conventions for the sign of the current contribution, this is the geometric action associated with the semidirect boundary structure
\begin{equation}
\Diff(S^1)\ltimes
\widetilde{\SL(2,\mathbb R)}
\label{eq:DiffKMsec8}
\end{equation}
as computed in \cite{Zuevsky:2018tah}, where $\widetilde{\operatorname{SL}(2,\R)}$ denotes the loop group. The dressing mechanism can also be restated as a Sugawara shift as shown in \cite{Gonzalez:2018enk}. In fact, after the field redefinition
\begin{eqnarray}
    \mathcal{L} \rightarrow \mathcal{L} + u(\tau), 
\end{eqnarray}
we get the action associated with coadjoint orbits of the aforementioned semi-direct product found in \cite{Zuevsky:2018tah, Gonzalez:2018enk, Chirco:2024ubu}. 

The important point is not the particular choice of the dressing function $b(\tau)$, but the general mechanism: the Schwarzian sector acquires a Kac--Moody extension once the second sector is reinterpreted in terms of loop-algebra currents  acted upon by the projective reparametrization mode of the first. This construction shows that specific configurations of three-dimensional gravity, after symmetry reduction, are effectively described by a one-dimensional action which reproduces the low energy dynamics of charged SYK-like models, providing an extended family of holographic relations beyond the standard JT/SYK one. 


 
\section{Discussion}
\label{discussion}
In this paper, we studied a symmetry-reduced version of three-dimensional
$AdS_3$ gravity formulated as an $SO(2,2)$ Chern--Simons
theory on a manifold with toroidal boundary. By imposing invariance of
the gauge connection along the non-contractible angular cycle, we showed
that each $\mathfrak{sl}(2,\mathbb{R})$ sector reduces to a
two-dimensional BF-like system,
\begin{equation}
    F_\perp=0,
    \qquad
    D_\perp\Phi=0,
\end{equation}
together with the one-dimensional boundary contribution
\begin{equation}
    S_{\mathrm{1D}}
    =
    -\frac{k_{\mathrm{CS}}}{2}
    \int_{\partial M_2}
    \operatorname{Tr}(\Phi A_\tau)\,d\tau.
\end{equation} 
A central point of the construction is that this boundary functional is
not introduced independently at the two-dimensional level, but is fixed
by the dimensional reduction of the parent 3D Chern--Simons action. 
We then focused on the first sector and analyzed the reduced theory on boundary subspace selected by boundary conditions in the form 
\begin{eqnarray}
A_{\tau} = \lambda'\Phi\,.
\end{eqnarray}
The equations of motion imply that the quadratic Casimir $   C=\operatorname{Tr}(\Phi^2)$
is conserved. After transforming the connection to
Drinfel'd--Sokolov gauge, the corresponding highest-weight component is
\begin{equation}
    \mathcal L(\tau)
    =
    -\frac{C}{2}\lambda'^{\,2}
    +
    \frac{1}{2}\{\lambda,\tau\}.
\end{equation}
The reduced functional can be related to the geometric action of \cite{Alekseev:1988ce,Alekseev:2018pbv, Nguyen:2021pdz}. 
The regularity
condition on the holonomy around the contractible Euclidean-time cycle
selects the constant Drinfel'd--Sokolov representative and 
 leads to the exceptional Virasoro
orbits relevant for the Schwarzian description of JT gravity.
It is useful in this respect to distinguish two kinds of boundary
restrictions. Conditions such as those above determine the
boundary subspace on which the variational principle is well posed.
Additional restrictions on the Lie-algebra components of the
connection, such as the dimensionally reduced form of
Brown--Henneaux-type conditions, specify the asymptotic sector in which
the Drinfel'd--Sokolov reduction is implemented. The Schwarzian mode
arises from the combination of these two ingredients.

Finally, we considered both
$\mathfrak{sl}(2,\mathbb{R})$ sectors of the $  \mathfrak{so}(2,2)$ factorization.
If they  are treated independently, the reduction produces
two decoupled and formally identical, one-dimensional systems. More interestingly, the second
sector may be organized as a loop-valued current sector acted upon by
the reparametrization mode of the first. With an appropriate dressing,
this gives a boundary action with the semidirect-product structure
\begin{equation}
    \Diff(S^1)\ltimes \widetilde{SL(2,\mathbb{R})},
\end{equation}
corresponding to a non trivial Kac--Moody extension of the Schwarzian theory. This construction realizes a three-dimensional gravitational embedding of
current-dressed Schwarzian actions of the type appearing in SYK-like
models with global symmetries. 

Establishing a  complete holographic interpretation of such an embedding within the JT/SYK duality deserves further investigation.

\section*{Acknowledgments}
The authors acknowledge support from the INFN Iniziativa Specifica GeoSymQFT and from the European COST Action CaLISTA CA21109. P.V. acknowledges support from the PNRR MUR Project No. CN 00000013-ICSC. L.V. thanks the group of G. Barnich and G. Compère for fruitful discussions and the Theoretical and Mathematical Physics of the ULB for hospitality while this work was being completed.

\bibliographystyle{JHEP}
\bibliography{DRnoncovariant}

\appendix

\section{\texorpdfstring{$\mathcal{L}(\tau)$} as a Schwarzian derivative}
\label{appendixA}
We can show that the flatness condition $A_{\tau}= g^{-1}\partial_{\tau}g$ together with $A_{\tau} = L_{+}+ \mathcal{L}(\tau)L_{-}$ implies that $\mathcal{L}(\tau)$ is the Schwarzian derivative of a function $F$ completely determined by $g$. Furthermore, $F$ is periodic. Given $g$ as 
\begin{eqnarray}
    g(\tau) = \begin{pmatrix}
        \psi_1& \psi_2\\ \phi_1 &\phi_2
    \end{pmatrix}, 
\end{eqnarray}
the equation we have to satisfy is 
\begin{eqnarray}
    g'(\tau) = \begin{pmatrix}
        \psi_1'& \psi'_2\\ \phi'_1 &\phi_2'
    \end{pmatrix} = g(\tau)A_{\tau} = \begin{pmatrix}
        \psi_1& \psi_2\\ \phi_1 &\phi_2
    \end{pmatrix}\begin{pmatrix}
        0& -\mathcal{L}(\tau)\\ 1 &0
    \end{pmatrix}
\end{eqnarray}
from which we get the set of equations
\begin{eqnarray}
    \psi'_1 = \psi_2, \quad \psi_2' = -\psi_1\mathcal{L}, \quad \phi_1'=\phi_2, \quad \phi'_2 = -\phi'_1\mathcal{L},
\end{eqnarray}
then $g(\tau)$ gets reduced to
\begin{eqnarray}
    g(\tau) = \begin{pmatrix}
        \psi_1& \psi_1'\\ \phi_1 &\phi_1'
    \end{pmatrix},
\end{eqnarray}
with
\begin{eqnarray}
    \psi_1'' + \mathcal{L}\psi_1 =0, \quad \phi_1'' + \mathcal{L}\phi_1 =0, \quad \psi_1\phi_1'-\psi_1'\phi_1 =1.
\end{eqnarray}
Consider now the ratio $F = \psi_1/\phi_1$, 
\begin{eqnarray}
    F(\tau)' = (\psi_1'\phi_1 -\psi_1\phi_1')/\phi_1^2 = 1/\phi_1^2,  
\end{eqnarray}
\begin{eqnarray}
    F'' = -2\phi_1'/\phi_1^3,  \quad F''/F' = -2\phi_1'/\phi_1 
\end{eqnarray}
and
\begin{eqnarray}
    (F''/F')' = F'''/F' - (F''/F')^2 = -2\phi_1''/\phi_1+2\phi_1'^2/\phi_1^2, 
\end{eqnarray}
thus
\begin{eqnarray}
\label{schwarzianL}
    \{F,\tau\} = -2\phi_1''/\phi_1+2\phi_1'^2/\phi_1^2 -1/2 (4 \phi_1'^2/\phi_1) = 2\mathcal{L}(\tau).
\end{eqnarray}
Moreover, $g(\tau+2\pi) = \pm g(\tau)$ automatically implies the periodicity of $F$
\begin{eqnarray}
    F(2\pi)-F(0)=0.   
\end{eqnarray}
The factor 2 is fundamental, since $\mathcal{L} = 1/2\{F,\tau\}_S$ is the only factor that allows the action to be rewritten as a total Schwarzian derivative.

\section{Covariant boundary conditions}
\label{app:covBC}
Here we first prove that the set of boundary conditions 
\begin{equation}
\label{bcondcov}
    \delta\Phi=0, \quad A_{\tau} = \lambda'(\tau)\Phi + u^{-1}\partial_{\tau}u,
\end{equation}
cures the boundary anomaly of the Chern-Simons variation \eqn{eq:deltaSredonshell}. Then we show that the general local form of $\Phi$ according to the boundary conditions and compatible with the equation of motion $D_A{_{\tau}}\Phi=0$ is
\begin{equation}\label{secondstat}
    \Phi = u^{-1}Tu,
\end{equation}
with $T$ an arbitrary generator in $\mathfrak{sl}(2,\mathbb{R})$.
The variation of the boundary gauge potential $A_{\tau}$, according to \eqn{bcondcov} is
\begin{eqnarray}
    \delta A_{\tau} = (\delta\lambda)' \Phi + \partial_{\tau}\epsilon + [u^{-1}\partial_{\tau}u,\epsilon]
\end{eqnarray}
where $\epsilon = u^{-1}\delta u$. Substituting into \eqn{eq:deltaSredonshell}, we get
    \begin{equation}
    \delta S_{CS}[A]_{\text{on shell}} =\frac{k}{4\pi}\int_{\partial M_3} \operatorname{Tr} \big(\Phi(\delta\lambda' \Phi + \partial_{\tau}\epsilon + [u^{-1}\partial_{\tau}u,\epsilon]) \big)\, d\tau  d\phi. .
\end{equation}
We know that $\operatorname{Tr}(\Phi^2)$ is constant on shell, therefore, the term in $\delta \lambda'$ is zero after integration by parts and upon implementing the $\phi$ independence of the fields. The terms in $\epsilon$ can be then rewritten as
    \begin{equation}
    \delta S_{CS}[A]_{\text{on shell}} =-\frac{k}{4\pi}\int_{\partial M_3} \operatorname{Tr}  \big(\epsilon(\Phi' + [u^{-1}\partial_{\tau}u, \Phi]) \big)\, d\tau  d\phi.
\end{equation}
Therefore, on using the  equation of motion of $\Phi$,  $D_{A_{\tau}}\Phi=0$,  with $A_\tau=\lambda'(\tau)\Phi + u^{-1}\partial_{\tau}u$, we obtain
\begin{equation}\label{eomphi}
    \Phi' + [u^{-1}\partial_{\tau}u, \Phi]=0,
\end{equation}
implying that the variation of the action  is zero on-shell. 
Let us consider the second statement, that is, Eq.\eqn{secondstat}. Since the equation of motion of $\Phi$ when restricted at the boundary takes the form \eqn{eomphi}, 
we can look for solutions of this equation. It can be easily checked that  that it can be rewritten in the form 
\begin{eqnarray}
    u^{-1}(u\Phi u^{-1})'u =0,
\end{eqnarray}
which is solved by 
\begin{equation}
    \Phi = u^{-1}Tu
\end{equation}
with $T$ a fixed element of the Lie algebra, {\it cvd}.
\end{document}